\documentclass[10pt, natbib]{article}
\usepackage{graphicx}    
\usepackage{amsmath}
\usepackage{amssymb}
\usepackage{verbatim}
\usepackage{multirow}
\usepackage{hyperref}
\usepackage{caption}
\usepackage{subcaption}
\usepackage[super]{natbib}
\usepackage[top=4.5cm, bottom=4.5cm, left=2.5cm, right=2.5cm, headsep=0pt, headheight=0pt, footskip=0pt, ]{geometry}
\usepackage{setspace}
\newcommand\BibTeX{{\rmfamily B\kern-.05em \textsc{i\kern-.025em b}\kern-.08em
T\kern-.1667em\lower.7ex\hbox{E}\kern-.125emX}}
\newcommand{\bs}{\ensuremath{\boldsymbol}}
\newcommand{\mb}{\ensuremath{\mathbf}}
\newcommand{\mbE}{\ensuremath{\mbox{E}}}
\newcommand{\rhoy}{\ensuremath{\rho_{ij} (y, \mb{X}_i )}}
\newcommand{\rhoyo}{\ensuremath{\rho_{ij} (y_0, \mb{X}_i )}}

\newcommand{\rhoyobs}{\ensuremath{\rho_{ij} (y, \mb{X}_{ij} )}}
\newcommand{\rhoyoobs}{\ensuremath{\rho_{ij} (y_0, \mb{X}_{ij} )}}
\newcommand{\odds}{\ensuremath{\mbox{odds}_S}}
\newcommand{\prob}{\ensuremath{\mbox{Pr}}}
\newcommand{\pione}{\ensuremath{\pi(1, \mb{X}_{1,i})}}
\newcommand{\pizero}{\ensuremath{\pi(0, \mb{X}_{1,i})}}

\newcommand{\pioneobs}{\ensuremath{\pi(1, \mb{X}_{1,ij})}}
\newcommand{\pizeroobs}{\ensuremath{\pi(0, \mb{X}_{1,ij})}}
\newcommand{\sumofsubj}{\ensuremath{\sum_{i} \sum_{j}}}
\newcommand{\lamy}{\ensuremath{\lambda_{P_{ij}}(y, \mb{X}_i)}}
\newcommand{\lamyo}{\ensuremath{\lambda_{P_{ij}}(y_0, \mb{X}_i)}}
\newcommand{\lamyobs}{\ensuremath{\lambda_{P_{ij}}(y, \mb{X}_{ij})}}
\newcommand{\lamyoobs}{\ensuremath{\lambda_{P_{ij}}(y_0, \mb{X}_{ij})}}
\newcommand{\wone}{\ensuremath{\mb{w}_{1,ij}}}
\newcommand{\wtwo}{\ensuremath{\mb{w}_{2,ij}}}
\newcommand{\ti}{\ensuremath{\mb{T}_i(\bs{\gamma})}}
\newcommand{\ui}{\ensuremath{\mb{U}_i(\bs{\gamma}, \bs{\beta})}}

\def\ci{\perp\!\!\!\perp}

\begin{document}


\title{Generalized Linear Models for Longitudinal Data with Biased Sampling Designs: A Sequential Offsetted Regressions Approach}

\author{Lee S. McDaniel$^{1}$, Jonathan S. Schildcrout$^{2}$, Enrique F. Schisterman$^{3}$, and Paul J. Rathouz$^{4}$ \\
$^{1}$Biostatistics Program, School of Public Health, \\Louisiana State University Health Sciences Center\\
$^2$Departments of Biostatistics and Anesthesiology, Vanderbilt University School of Medicine\\
$^{3}${\it Eunice Kennedy Shriver} National Institute of Child Health and Human Development \\
$^{4}$Department of Biostatistics and Medical Informatics,\\ University of Wisconsin, Madison}




\begin{abstract}
Biased sampling designs can be highly efficient when studying rare (binary) or low variability (continuous) endpoints.  We consider longitudinal data settings in which the probability of being sampled depends on a repeatedly measured response through an outcome-related, auxiliary variable.  Such auxiliary variable- or outcome-dependent sampling improves observed response and possibly exposure variability over random sampling, {even though} the auxiliary variable is not of scientific interest.  {For analysis,} we propose a generalized linear model based approach using a sequence of two offsetted regressions.  The first estimates the relationship of the auxiliary variable to response and covariate data using an offsetted logistic regression model.  The offset hinges on the (assumed) known ratio of sampling probabilities for different values of the auxiliary variable.  Results from the auxiliary model are used to estimate observation-specific probabilities of being sampled conditional on the response and covariates, and these probabilities are then used to account for bias in the second, target population model. We provide asymptotic standard errors accounting for uncertainty in the estimation of the auxiliary model, and perform simulation studies demonstrating substantial bias reduction, correct coverage probability, and improved design efficiency over simple random sampling designs. We illustrate the approaches with two examples.
\end{abstract}


\maketitle

\section{Introduction}

{Outcome dependent sampling designs for epidemiologic and health outcomes
  studies are ubiquitous in medical 
and public health research.  By enriching the sample with relatively informative data units, they can be 
highly efficient when compared to simple random sampling designs.   We consider a
class of such designs specifically geared towards longitudinal response data.  These designs sample 
on an auxiliary variable that is related, but not equal, to the response vector.  Even though the auxiliary variable
 is used to enrich the sample with relatively informative subjects (under subject-level sampling) and/or timepoints (under observation-level
sampling), in the framework considered here, it is not of scientific interest.  The potential for efficiency gains improves to the extent that the auxiliary variable is, in fact, related 
to the response vector.}

{A}ssume a target population wherein each subject $i$ has
attributes $( \mb{Y}_i , \mb{t}_i , \mb{X}_i, \mb{Z}_i)$, with $\mb{Y}_i = ( Y_{i1},
\ldots, Y_{i n_i})'$ a longitudinal series of continuous or discrete
outcomes, $\mb{X}_i = (\mb{X}_{i1}, \ldots, \mb{X}_{i n_i})'$ an $n_i
\times p$ matrix of covariates predicting $\mb{Y}_i$, and $\mb{t}_i =
(t_{i1}, \ldots, t_{i n_i} )'$ a vector of potential observation times.
Interest lies in the marginal regression model $\mbE(\mb{Y}_{ij} |
\mb{X}_{ij})$ parameterized by $\bs{\beta}$, and any effects of $t_{ij}$ are
encoded in~$\mathbf{X}_{ij}.$ For sampling, assume a binary, auxiliary $Z_{ij}$ and, optionally,
$\mathbf{X}_{1,ij}$, contained in $\mb{X}_{ij}$, available on all
members---and at all time points---of the target population.

\emph{For observation-level sampling}, assume the probability of an observation
being sampled
at any time point $t_{ij}$ is based only on $Z_{ij}$ and, optionally, on
$\mathbf{X}_{1,ij}$. The stratifying variable $\mathbf{X}_{1,ij}$ may be
a confounder or a key predictor of interest.  Full data $({Y}_{ij }, \mb{X}_{ij})$ are only
observed at sampled time points.   Such a design was
implemented in the Biocycle Study, which examined the
association among oxidative stress levels, antioxidant levels, dietary
intake, and reproductive hormone concentrations over the course of the
menstrual cycle in approximately 250 participants \citep{pmid19159403,
  pmid18974081, schildcrout2012outcome}. 
 To improve estimation efficiency, the Biocycle Study sought to over-sample
relatively informative peri-ovulation days, where hormone concentrations change rapidly. 
To this end, participants were provided at-home, urine-based fertility tests to be used daily until 
the day of peak fertility.  On the day of peak fertility and the two subsequent days, 
participants visited study clinics for peri-ovulation blood draws.  Blood draws also
occurred on five other days during the cycle.  In this study, the result of
the at-home fertility test is an auxiliary variable $Z_{ij}$ on which time-point specific
sampling is based, and the probability of being sampled is high when $Z_{ij}=1$ (peri-ovulation days) and is low
when $Z_{ij}=0$ (other days during the cycle).

\emph{In the case of subject-level sampling}, assume the probability of being sampled
depends on a time-constant binary variable, $Z_i$, that is related to
$\mb{Y}_i$ and possibly some subset $\mb{X}_{1,i}$ of $\mb{X}_i.$  
This formal framework is motivated by a natural history study of attention deficit
hyperactivity disorder (ADHD) \citep{lahey1998validity, hartung2002sex, S:Schildcrout+Rathouz:2010}. 
The ADHD Study sampled 138
children based on a clinical referral at baseline ($t_{i1}$) due to parent or teacher suspicion of
ADHD, and 117  demographically and socioeconomically similar
non-referred children from the same community. The binary auxiliary
variable $Z_i$, indicating whether or not a child is referred,
is related to downstream psychological outcomes $Y_{ij}$ such as
level of ADHD symptoms or time-specific diagnoses of ADHD.  The sample was
enriched because informative children with $Z_i=1$ (high risk of ADHD symptoms) were sampled 
with high probability while less informative children with $Z_i=0$ (low risk of ADHD symptoms) were sampled 
with a much lower probability.

A semiparametric, marginal model estimation strategy has been explored for binary response data in
the subject-level sampling case by Schildcrout and Rathouz~\cite{S:Schildcrout+Rathouz:2010} and
in the observation-level sampling case by Schildcrout~et~al.\cite{schildcrout2012outcome}.
Although those papers also provide a review of the prior literature on this and
related problems, recently Neuhaus~et~al.\ \cite{BIOM:BIOM12108} has employed
a parametric generalized linear mixed model approach to estimate conditional model
parameters in the setting of subject-level sampling.
The present work describes a general framework for conducting auxiliary variable
dependent sampling designs and associated semiparametric, marginal model analyses 
that encompasses both subject-level and observation-level sampling and that applies to 
all members of the exponential family (i.e., all generalized linear models).  The extension from 
analyses of binary data to count and continuous data is non-trivial since, for example, study 
design considerations (i.e., when the study design is useful) and the potential for over- or 
under-dispersion (not a challenge with binary data) depend on the response 
distribution.  Analytical tools for these biased designs are generalized from binary logistic regression
to all generalized linear models using the odds model representation of  
exponential family models described in Rathouz and Gao \cite{RathouzGao}.  Though
originally developed under random sampling scenarios, we show that it can also be used
for analyses of biased samples.  Finally, an R package, SOR, is available on CRAN (https://cran.r-project.org/) to conduct
analyses described herein.

In Section 2 we describe the model of interest, observation-level and subject-level sampling designs, and modeling assumptions that must be made for valid inferences in each design.  Section~3 describes our general approach for estimation and inference.  In
Section~4 we investigate, via simulation, operating
characteristics of the proposed estimators and an inverse probability weighted (IPW) estimation
approach, looking specifically at bias, coverage probability, and relative efficiency of the designs. 
We apply the analytical techniques to the ADHD and Biocycle studies in Section~5, and
discuss our findings in Section~6.

\section{Study Designs and Analysis Models}

\subsection{Response Model in the Target Population}

Interest lies in the marginal or ``population averaged'' 
mean regression model,
\begin{equation}
\label{population mean}
\mu_{P_{ij}} = \mbox{E}(Y_{ij} | \mb{X}_{ij}) = g^{-1} (\beta_0 + \mb{X}_{ij}' \bs{\beta}_1)
\end{equation}
(subscript $P$ for target population and, below, $S$ for sample), where $g(\cdot )$
is a link function and $\bs{\beta}=(\beta_0,\bs{\beta}_1)$, or elements thereof, is 
the parameter of interest.
Letting $F_P ( y | \mb{X}_{ij} )$ be the conditional (on $\mb{X}_{ij}$)  distribution function
for the response $Y_{ij}$ in the target population, marginally over the
other $Y_{ij'}$'s, we assume that $dF_P (y |
\mb{X}_{ij})$ belongs to an exponential family with
\begin{align}
\label{pop density}
dF_P (y | \mb{X}_{ij}) \equiv  f(y | \mb{X}_{ij})  
= & \exp \left\{ [\theta_{ij} y - b ( \theta_{ij}
  )]/\phi  + c(y; \phi) \right\}\ .
\end{align}
Specifying  $\mbE(Y_{ij} | \mb{X}_{ij} ) = b' ( \theta_{ij}) = g^{-1}(\eta_{ij})$ and 
$\eta_{ij} = \beta_0 + \mb{X}_{ij}' \bs{\beta}_1,$
(\ref{pop density}) becomes a generalized linear model with canonical
parameter $\theta_{ij}$, cumulant function $b(\theta_{ij})$, and dispersion parameter~$\phi$.  

For example, if $Y_{ij}$ is continuous---e.g.,
the logarithm of measured luteinizing hormone in the BioCycle Study---we
would typically use a linear regression model with Gaussian errors. If
$Y_{ij}$ is a count---e.g., the number of hyperactivity symptoms in the
ADHD Study---we might use a log-linear regression with Poisson errors.

Towards developing our analysis approach,  it is useful to observe that from (\ref{pop density}), we can express $dF_p(\cdot)$ in terms of the
population odds
\begin{align*}
\mbox{odds}_P (y|\mb{X}_{ij}) \equiv  \frac{dF_P (y | \mb{X}_{ij})}{dF_P
  (y_0 | \mb{X}_{ij} ) } 
= \mbox{exp}\{ \theta_{ij} (y - y_0 )/ \phi + c(y;
\phi) - c(y_0; \phi)\},
\end{align*}
where $y_0$ is an arbitrary reference value of the response distribution
(e.g., the population median) such that $dF_P (y_0 | \mb{X}_{ij} = \mb{x})
> 0 \ \forall \mb{x}$.  Whereas this is a common representation in the binary case, with $y_0=0$, the odds
model representation is generalizable to any other outcome distribution in
the exponential family (e.g., Rathouz and Gao\cite{RathouzGao}).

\subsection{Response Model in Sampled Data}

Let $S_{ij}=1$ if subject $i$ is observed at the $j^{th}$ timepoint and
$0$ otherwise, noting that $S_{ij}=S_i$ for all $j$ in subject-level
sampling designs.  Applying Bayes' Theorem, and irrespective of the
design, define the conditional (on $S_{ij}=1$) density for response
$Y_{ij}$ in the sampled data as
\begin{align}
\label{sample density}
dF_{S} (y | \mb{X}_{ij} )  \equiv  f(y | \mb{X}_{ij}, S_{ij}=1) 
 =  \frac{dF_P (y | \mb{X}_{ij} ) \times
  \mbox{Pr}(S_{ij} =1 | Y_{ij} = y, \mb{X}_{ij} )}{\mbox{Pr} (S_{ij} =1 |
  \mb{X}_{ij} )} \ .
\end{align}
From~(\ref{sample density}), it can now be shown that the odds model in the sample is,
\begin{align} \label{sample odds}
\odds (y | \mb{X}_{ij}) \equiv & \frac{ dF_{S} (y | \mb{X}_{ij} )}{ dF_{S} (y_0 | \mb{X}_{ij} )}
=  \frac{\mbox{odds}_P (y | \mb{X}_{ij})\rhoyobs}{\rhoyoobs}   
=  \exp \left\{ \frac{\theta_{ij} (y - y_0)}{ \phi}  + c^*(y; \phi)
\right\} \ ,
\end{align}
where $\rhoyobs \equiv \mbox{Pr}(S_{ij} = 1 | Y_{ij} = y, \mb{X}_{ij})$, and
\begin{equation}
\label{cstar}
c^* (y; \phi) = c(y; \phi) - c(y_0; \phi) + \log \left\{ \frac{\rhoyobs}{\rhoyoobs }\right\}
\ .
\end{equation}
We thus obtain
\begin{equation}
\label{sample density two}
dF_S (y | \mb{X}_{ij} ) = \exp \left\{ [ \theta_{ij}y  - b^*
  (\theta_{ij})]/\phi + c^* (y; \phi) \right\}\ ,
\end{equation}
where
$$
b^* (\theta_{ij}) = \theta_{ij} y_0 + \log \int_{y} \odds (y | \mb{X}_{ij}
) dy \ .
$$
In the following sections, 
we pursue estimation of the sampling ratio
$\rhoyobs/\rhoyoobs,$
which is essential to~(\ref{sample density two}), and show how it is used
for inferences on  mean model parameter $\bs{\beta}.$

An important consequence of this development is that,
conditional on the subject being sampled, the model for $Y_{ij}$ is still
in the exponential family of models, with the same canonical parameter
$\theta_{ij}$.  As long as the sampling ratio $\rhoyobs/\rhoyoobs$ does
not depend on $\theta_{ij}$ (or $\bs{\beta}),$ only the reference
distribution $\exp\{c^*(y;\phi)\}$ (properly normalized) changes, as in
Equation~(\ref{cstar}).

\subsection{Longitudinal Data with Observation-Level Sampling}

Assume sampling is conducted at the level of observations within subjects, based on a
time-varying auxiliary variable $Z_{ij}$, or jointly on $Z_{ij}$ and 
$\mathbf{X}_{1, ij}$. 
Let $\mb{S}_i=\{S_{i1}, \dots S_{in_i}\}$ contain time-specific sampling
indicators $S_{ij}$ over times $t_{ij}$, $j \in \{1, \ldots, n_i\}$.  For example, in the BioCycle
study, $Z_{ij} = 1$ if the subject's at-home fertility test indicates peak fertility
on day $j$ or within the last two days. In that case, $S_{ij}=1$
with probability 1. On all other days, $Z_{ij} = 0$, and $S_{ij}=1$ with a
relatively low but non-zero probability.
Assuming sampling is based exclusively on $(Z_{ij}, \mathbf{X}_{1, ij})$, then
\begin{align}
\label{obs sampling assumption}
\mbox{Pr}(S_{ij} =1 | Z_{ij}=z , Y_{ij}, \mb{X}_{ij}) = 
\mbox{Pr}(S_{ij} = 1| Z_{ij}=z, \mb{X}_{1, ij})  
{\equiv} 
\pi(z, \mb{X}_{1, ij}), \ 
\end{align}

\noindent which, as part of the study design, is known to and can therefore be 
specified by the investigator.

In overview, analysis proceeds by using the known sampling
ratio $\pioneobs/$ $\pizeroobs$ to estimate $\rhoyobs / \rhoyoobs,$
which in turn is used for inferences regarding $\bs{\beta}$.  
To obtain $\rhoyobs / \rhoyoobs,$ 
we use an intermediary, auxiliary variable model $\lambda_{P_{ij}} (y,
\mb{X}_{ij}) \equiv \prob (Z_{ij} =1 | Y_{ij} = y, \mb{X}_{ij} )$ relating
$Z_{ij}$ to $(Y_{ij}, \mb{X}_{ij})$ in the target population. 
Conditioning on sampling $S_{ij}=1$, 
we estimate 
$\lambda_{P_{ij}} (y, \mb{X}_{ij})$
by estimating
$\lambda_{S_{ij}}(y, \mb{X}_{ij}) \equiv \prob(Z_{ij} = 1 | Y_{ij} = y,
S_{ij} = 1, \mb{X}_{ij})$ and by then exploiting the 
known relationship between $\lambda_{P_{ij}} (y, \mb{X}_{ij})$ and
$\lambda_{S_{ij}} (y, \mb{X}_{ij})$ under our design.  This intermediary model technique follows from Lee et al.\cite{SIM:SIM557}, Neuhaus et al.\cite{BIOM:BIOM450}, and Schildcrout and Rathouz\cite{S:Schildcrout+Rathouz:2010}.

To pursue this program, first specify a model for $\lamyobs$.  Whereas any binary data
regression model, 
with its attendant modeling assumptions, is a legitimate approach, logistic regression simplifies subsequent development. Posit
\begin{equation}
\label{population lambda}
\lamyobs = \mbox{logit}^{-1} \{ \wone' \gamma_1 + h(y) \times \wtwo'
\gamma_2\} ,
\end{equation}
where $\wone$ and $\wtwo$ are functions of $\mb{X}_{ij}.$
Here, 
$h(\cdot)$ is a user-specified function introduced to allow dependence of
$\lamyobs$ on $y$; it should be chosen to reflect the relationship between the auxiliary variable and $Y_{ij}$.
The key here is that the specified forms of $h(\cdot),$ $\wone,$ and $\wtwo$ contain enough information so that
\begin{equation}
\label{richness of lambda model}
\prob (Z_{ij} =1 | Y_{ij} = y, \wone, \wtwo) = \prob (Z_{ij} =1 | Y_{ij} =
y, \mb{X}_{ij})\ ,
\end{equation}
at least to a good approximation.
See Scihldcrout et al.\cite{schildcrout2012outcome} for a discussion on the specification of this model in the observation-level sampling case.

We then note the connection between the unobserved population
and the pseudo-population represented by the sample
with respect to $\lamyobs$. By Bayes' Theorem,
\begin{equation}
\label{lambda odds}
\frac{\lambda_{S_{ij}} (y, \mb{X}_{ij})}{1-\lambda_{S_{ij}}(y,
  \mb{X}_{ij})} = \frac{\lamyobs}{1-\lamyobs} \cdot
\frac{\pioneobs}{\pizeroobs}\ .
\end{equation}
Equations (\ref{population lambda}) and (\ref{lambda odds}) then yield a
model for $\lambda_{S_{ij}}$, 
\begin{align}
 \label{sample lambda}
\lambda_{S_{ij}}(y, \mb{X}_{ij}) = 
 \mbox{logit}^{-1} \left[ \wone'
  \gamma_1 + h(y)\times \wtwo' \gamma_2 + \log \left\{ \frac{\pioneobs}{\pizeroobs}
  \right\} \right] \ ,
\end{align}
which is fitted to the sample data for estimation of~$\bs{\gamma} \equiv (\gamma_1, \gamma_2)$. 
Finally, estimation of~$\bs{\beta}$ obtains from the density $dF_S(\cdot)$
given in~(\ref{sample density two}).    Using $dF_S(\cdot)$ involves log
transforming
the estimated sampling ratio
\begin{equation}
\label{rho ratio}
 \frac{\rhoyobs}{\rhoyoobs} =  \frac{ 1   - \lamyobs +  \frac{\pioneobs}{\pizeroobs} \lamyobs}
{1 - \lamyoobs +  \frac{\pioneobs}{\pizeroobs} \lamyoobs}\ ,
\end{equation}
and plugging it into~(\ref{cstar}).

\subsection{Longitudinal Data Following Subject-Level Sampling}

We address subject-level sampling as a special case of
observation-level sampling. Let $S_{ij}=S_i$ be an indicator which is $1$
for all $j = 1, \ldots, n_i$ if the $i$th subject is sampled, and 0
otherwise. 
We consider a design in which $S_{i}$ is related to $\mb{Y}_i$ and possibly $\mb{X}_i$.
Assume sampling depends on $(\mb{Y}_i, \mb{X}_i )$ only through binary
$Z_i$ and possibly through some subset $\mb{X}_{1,i}$ of $\mb{X}_i,$ and that sampling is independent across subjects.     Formally,
\begin{equation}
\label{sampling assumption}
S_{i} \ci (\mb{Y}_i , \mb{X}_i ) | (Z_i, \mb{X}_{1,i})\ .
\end{equation}
In the ADHD study, $Z_i = 1$ if the subject is referred by a parent or
teacher.
Leveraging the sampling probability at the observation level, $\rhoyobs \equiv
\mbox{Pr}(S_{i} = 1 | Y_{ij} = y, \mb{X}_{ij})$, the remainder of the
development is tightly analogous to that for observation-level
sampling. Note that $\rhoyobs$ is the probability of being sampled
conditional only on the $j$th response $Y_{ij}$ and $j$th covariate vector
$\mathbf{X}_{ij}$ (versus the entire vector $\mb{Y}_i$ and matrix
$\mb{X}_i$). 

\section{Estimation and Inference}

\subsection{Estimation: Model of Interest and Auxiliary Model}

We present a procedure for coefficient estimation and inference for the
case of observation-level sampling. This approach also applies to
subject-level sampling by setting $S_i=S_{ij}$. In addition, a more statistically efficient
modification is available for a special case of subject-level sampling;
this is pursued in Section~3.2.

To estimate coefficients $\bs{\gamma}$ in equation (\ref{sample lambda}), we solve the standard logistic regression score equation $\sum_{i} \mb{T}_i (\bs{\gamma}) = \bs{0}$, where
\begin{equation}
\label{lambda score}
\mb{T}_i (\bs{\gamma}) = \sum_{j=1}^{n_i} \begin{pmatrix} \wone \\
  h(Y_{ij}) \times \wtwo \end{pmatrix} ( Z_{ij} - \lambda_{S_{ij}} )\ .
\end{equation}
With estimates $\widehat{\bs{\gamma}}$, 
we can estimate $\lambda_{S_{ij}}(y,
\mb{X}_{ij})$, which permits estimation of 
$\lambda_{P_{ij}}(y, \mb{X}_{ij})$ and $\rhoyobs / \rhoyoobs$ 
by plugging estimates  $\widehat{\bs{\gamma}}$ into
equations (\ref{population lambda}) and (\ref{rho ratio}).  This leads to the biased-sample mean model through equation (\ref{sample odds}):
\begin{equation}
\label{sample mean}
\mu_{S_{ij}} = \frac{\int_y y \times \mbox{odds}_S (y | \mb{X}_{ij} )
  dy}{\int_y \mbox{odds}_S (y | \mb{X}_{ij} ) dy} \ ,
\end{equation}
forming the basis for 
estimation of and inferences on $\bs{\beta}$. 

We pursue a generalized estimating equations (GEE) approach to inferences on~$\bs{\beta}$ by solving $\sum_i \mb{U}_i (\bs{\beta},
\widehat{\bs{\gamma}}) = 0$ for $\bs{\beta}.$  Here,
\begin{equation}
\label{beta equation}
\mb{U}_i (\bs{\beta}, \bs{\gamma}) = \mb{D}_i' \mb{V}_i^{-1} ( \mb{Y}_i -
\bs{\mu}_{S_i} )\ ,
\end{equation}
$\bs{\mu}_{S_i} = ( \mu_{S_{i1}}, \ldots, \mu_{S_{in_i}} )'$ (given in \ref{sample mean}), $\mb{D}_i' = (1_{n_i} , \mb{X}_i )' \mb{A}_i$,
\begin{align}
\label{var equation}
\mb{A}_i =  \mbox{diag}  \left\{  \frac{\int_y y^2 \times \mbox{odds}_S (y | \mb{X}_{ij} ) dy}{\int_y \mbox{odds}_S (y | \mb{X}_{ij} ) dy} 
- \left( \frac{\int_y y \times \mbox{odds}_S (y | \mb{X}_{ij} ) dy}{\int_y \mbox{odds}_S (y | \mb{X}_{ij} ) dy} \right)^2 \right\}_{j=1}^{n_i},
\end{align}
$\mb{V}_i = \mb{A}_i / \phi$, and noting that the $j^{th}$ diagonal element of $\mb{A}_i$ is the canonical variance in the sample.

We note that the foregoing GEE
procedure uses the independence working correlation structure.  This is
necessary because conditioning on $S_{ij}$ can lead to violation of 
the no interference  or full covariate conditional mean assumption that
$E(Y_{ij} \mid \mb{X}_{ij},S_{ij}=1)=E(Y_{ij} \mid \mb{X}_{i},S_{ij}=1)$ which, 
in turn, could introduce bias in
estimates of time-varying covariate parameters in~$\bs{\beta}$
\citep{sullivan1994cautionary,schildcrout2005regression}. 

\subsection{Special case of subject-level sampling}

The foregoing material in Sections 2 and 3, with the exception of
Section~2.4, applies to both observation- and subject-level sampling.
This obtains by letting $S_{ij}=S_i$ and $Z_{ij} = Z_i$ for all time
points $j=1, \cdots, n_i$.  
That is, $Z_i$ is repeated for the $i$th
subject $n_i$ times in~(\ref{lambda score}).  
In spite of this, estimation
of $\bs{\gamma}$ via (\ref{lambda score}) remains consistent.

There is, however, a special case of subject-level sampling wherein a more efficient procedure is possible. First, in the population, suppose the
``no interference'' assumption that $\mbE(Y_{ij} | \mb{X}_i) = \mbE(Y_{ij} | \mb{X}_{ij})$ holds, i.e., that information contained in $\mb{X}_{ij}$ is rich enough that predictors in $\mb{X}_i$ provide no additional predictive value beyond that from~$\mb{X}_{ij}$. In this case,
(\ref{population mean}) becomes a model for $\mbE(Y_{ij} | \mb{X}_i).$

Second, replace the observation-level sampling probability $\rhoyobs$ with
a special subject-level version $\rhoy= \mbox{Pr}(S_{ij} = 1 | Y_{ij} = y,
\mb{X}_{i}).$ Note that $\rhoy$ is the probability of being sampled
conditional on the entire design matrix $\mb{X}_i$, as opposed to
only the covariates at time $j$, but only on the $j$th response, $Y_{ij}.$

 Similarly, parallel to the development in Section~2.3, redefine $\lambda_{P_{ij}} (y, \mb{X}_{i}) \equiv \prob (Z_{i} =1 | Y_{ij} = y, \mb{X}_{i} )$ and $\lambda_{S_{ij}}(y, \mb{X}_{i})
\equiv \prob(Z_{i} = 1 | Y_{ij} = y, S_{i} = 1, \mb{X}_{i}).$ Then, because $Z_i$ is a scalar and $\pi(z,\mathbf{X}_{1,i})$ depends on the entire covariate matrix $\mathbf{X}_i$ through $\mathbf{X}_{1,i}$, we can estimate the sampling ratio 
\begin{equation}
\label{samp rho ratio}
\frac{\rhoy}{\rhoyo} = \frac{ 1 - \lamy + \frac{\pione}{\pizero}\lamy}{1 - \lamyo + \frac{\pione}{\pizero} \lamyo}\ ,
 \end{equation}
 which is analogous to (\ref{rho ratio}).

The remaining development parallels Section~2.2. It yields the conditional density $dF_{S} (y | \mb{X}_{ij} )$ for response $Y_{ij}$, satisfying the ``no interference'' assumption in the sample, i.e., $dF_{S} (y | \mb{X}_{ij} ) = dF_{S} (y | \mb{X}_{i} ).$ Owing to
this last property, statistical efficiency can be improved by using a working correlation model \begin{equation} \label{working corr} c_{S_{ijk}} = \mbox{corr}( Y_{ij}, Y_{ik} | \mb{X}_i , S_i =1 ; \bs{\alpha} )\ , \end{equation} as in a typical GEE setup. The resulting estimates
for $\bs{\alpha}$ are, however, for the correlation conditional on being sampled. Therefore it is not appropriate to use $\bs{\alpha}$ estimates to make inferences about the target population. If the working correlation model is a reasonable approximation to the true correlation
(given the sampling plan), then it should increase the efficiency of the estimates of $\bs{\beta}$ compared to estimation under the independence working correlation model \citep{liang1992multivariate, fitzmaurice1995caveat, mancl1996efficiency, schildcrout2005regression}. In
order to take advantage of the benefits of the working correlation model, we use $\mb{V}_i = \mb{A}_i^{1/2} \mb{C}_i \mb{A}_i^{1/2} / \phi$, where $\mb{C}_i$ is the $n_i \times n_i$ matrix with element $(j, k)$ given by (\ref{working corr}).

\subsection{Dispersion Parameter Estimation}

We turn now to the estimation of $\phi$, the dispersion parameter.  Assuming working independence across observations and subjects along with the model in (\ref{sample density}), we
calculate a likelihood-based score equation for $\phi$. Derivations in Appendix~A yield 
\begin{figure*}
\begin{align} \nonumber  \frac{d \log L}{d \phi}= &\frac{1}{\phi^2} \sumofsubj \left\{ - \theta_{ij} Y_{ij} + b (\theta_{ij}) \right\} + \sumofsubj c' (Y_{ij}; \phi) \\ \label{dispersion score}
& - \sum_{i} \sum_{j} \left\{\frac{1}{\int_{y}\rhoyobs dF_P(y | \mb{X}_{ij}) }    \int_{y} \left[ \frac{ - \theta_{ij}y + b (\theta_{ij})}{\phi^2} +
c'(y; \phi) \right]\rhoyobs dF_P(y | \mb{X}_{ij}) \right\} . 
\end{align}
\end{figure*}
As in a standard GEE implementation, for fixed $\bs{\beta}$, we set $d \log L / d\phi = 0$ and solve this equation numerically. After
solving for $\phi$, we return to estimation of $\bs{\beta}$ and iterate between the two until convergence.

\subsection{Empirical Standard Errors}

When the dispersion parameter is known, standard errors obtain via the sandwich estimator for variance. We proceed by viewing $ ( \widehat{\bs{\gamma}}, \widehat{\bs{\beta}})$ as the solution to the ``stacked'' estimating equation
\begin{equation} 
\sum_i \begin{pmatrix} \ti \\ \ui \end{pmatrix} = \mb{0} \ . 
\end{equation} 
Then the asymptotic variance of $( \widehat{\bs{\gamma}}',
\widehat{\bs{\beta}}')' $ is given by 
\begin{equation} \label{stderrs} \widehat{\mbox{AVar}} ( \widehat{\bs{\gamma}}' ,
\widehat{\bs{\beta}}' )' = \widehat{\mb{I}}^{-1} \widehat{\mb{Q}}
\widehat{\mb{I}}^{-1'}, 
\end{equation} 
where 
\begin{equation} \label{bread} \mb{Q} = \sum_{i} \begin{pmatrix}\ti \\
    \ui \end{pmatrix}^{\otimes 2}\ , \mbox{\ and} \quad \mb{I} = \begin{pmatrix} \mb{I}_{TT} & \mb{0} \\
\mb{I}_{UT} & \mb{I}_{UU} \end{pmatrix}\ . 
\end{equation} 
In (\ref{bread}),
\begin{align*}
 \mb{I}_{TT} = & \sum_i \mbox{E}\left( - \frac{\partial \mb{T}_i }{\partial \bs{\gamma}' } \right) \\
= & \sum_i \sum_{j=1}^{n_i} \begin{pmatrix} \wone \\ h(Y_{ij}) \times \wtwo
\end{pmatrix}^{\otimes 2} \{ \lambda_{S_{ij}} (1 - \lambda_{S_{ij}} )\}, \end{align*}
 the upper right hand quadrant of $\mb{I}$ is $\mb{0}$ because $\mbox{E}( - \partial \mb{T}_i / \partial \bs{\beta}' ) = \mb{0}$, 
\begin{equation*} \mb{I}_{UU} = \sum_i \mbox{E} \left( -
\frac{\partial \mb{U}_i}{\partial \bs{\beta}'} \right) = \sum_i \mb{D}_i'\mb{V}_i^{-1} \mb{D}_i,
 \end{equation*} 
and
 \begin{equation*} \mb{I}_{UT} = \sum_i \mbox{E} \left( - \frac{\partial \mb{U}_i }{\partial \bs{\gamma}' } \right) = \sum_i \mb{D}_i' \mb{V}_i^{-1} \left(
\frac{\partial \bs{\mu}_{S_i}}{\partial \bs{\gamma}' } \right), 
\end{equation*} 
where $\partial \bs{\mu}_{S_i}/\partial \bs{\gamma}'$ is given in Appendix~B. If the dispersion parameter $\phi$ must be estimated, then these uncertainty estimates may not be exactly correct, as they
do not take into account variability due to estimation of~$\phi$. Our simulation studies have shown,
however, that inferences on $\bs{\beta}$ are very robust to having
estimated~$\phi$, rather than it being known.

\subsection{An Alternative Approach: Inverse Probability Weighting}

It is common to use inverse probability weighting (IPW) to adjust for non-representative samples in missing data and survey sampling scenarios.  In its simplest form, IPW involves re-weighting observed subjects and/or observations by the inverse probability of being observed \citep{robins1994, robins1995analysis} so that the sample is representative of the target population.  Such IPW is easily applied under subject-level or observation-level sampling using GEE with inverse probability weight $1 / \pi(z, \mb{X}_{1,ij})$.  Although other more complex and efficient forms of IPW exist, we only investigate this simple IPW because it is very commonly applied and because we think of it as the only form of IPW that is more straightforward to implement than the proposed SOR procedure using our SOR package (available at https://cran.r-project.org/).

\section{Simulation}

\subsection{Subject-Level Sampling}

For each subject, we generate data with $n_i$ observations, where $n_i$ is $3, \ldots, 8$ with equal probability, and time $t_{ij}$ ranges from $0$ to $n_{i}-1$. Response $Y_{ij}$ is marginally Poisson with mean
\begin{equation} \label{poissimmean} \mu_{P_{ij}} = \exp\left( \beta_0 + \beta_{x_1} x_{1i} + \beta_t t_{ij} + \beta_{t, x_{1}} t_{ij} x_{1i} \right) \ . \end{equation} Here, $x_{1i}$ is a time-invariant binary covariate with $\prob(x_{1i}=1) = 0.15$ or 0.5. Responses follow an
exchangeable correlation structure with correlation parameter $\alpha = 0.5$, with data generated as in Yahav and Shmeuli\cite{yahav2012}.  Coefficient values are enumerated in Table~{\ref{poissonsim}}, and with these values, the model induces a marginal mean at $t_{i1}$ of $\mbE(Y_{i1}) =
0.25$ that drops by 9.5\%\ ($e^{-0.1}$) per unit of time if $x_{1i} = 0,$ and remains constant if $x_{i1} =1$.

A binary sampling covariate $Z_i$ is generated for each subject in the population using $\prob(Z_i=1|\mathbf{Y}_i,\mathbf{X}_i) = \mbox{logit}^{-1}(\gamma_0 + \gamma_1 I_{ \left\{ Y_{i1} >=1 \right\}} )$, where $\gamma_0 = -3.15$ and $\gamma_1 = 6.3$. We consider four design
strategies: simple random sampling [SRS; $\pi(z, x_{1i})=\pi$], covariate or exposure dependent sampling [ES; $\pi(z, x_{1i})=\pi(x_{1i})$], auxiliary variable dependent sampling [AVS; $\pi(z, x_{1i})=\pi(z)$], and exposure and auxiliary variable dependent sampling [EAVS; $\pi(z,
x_{1i})=\pi(z,x_{1i})$]. In the last three designs, the intent is to maximize observed (i.e., sampled) variability in sampling variable $Z_i$ and/or $X_{1i}$. To this end, we sample equal numbers in each stratum defined by $X_{1i}$, $Z_i$, and $(Z_i,X_{1i})$ under ES, AVS, and
EAVS, respectively. In all cases, we sample from a population of $N=100,000$ subjects.  Each subject has a probability of being sampled that leads to a total average sample size of 500.  Consequently, for our simulations, the sampling probabilities are given by:
\begin{eqnarray*}
\lefteqn{\text{SRS}:  \pi(z, x_{1i}) =  \pi =  \frac{500}{N} } \\
\lefteqn{\text{ES}:  \pi(z,x_{1i}) =  \pi(x_{1i})  =  500 \left(  \frac{x_{1i}}{\sum_{i=1}^N x_{1i}} +  \frac{(1-x_{1i})}{\sum_{i=1}^N (1-x_{1i})} \right) }\\
\lefteqn{\text{AVS}:  \pi(z, x_{1i}) =   \pi(z) = 500 \left(  \frac{z}{\sum_{i=1}^N z_i} +  \frac{(1-z)}{\sum_{i=1}^N (1-z_i)} \right) } \\
\lefteqn{\text{EAVS}:  \pi(z, x_{1i}) =  \pi(z, x_{1i}) =} \\ & & 500 \left(  \frac{z x_{1i}}{\sum_{i=1}^N z_i x_{1i}} +  \frac{z (1-x_{1i})}{\sum_{i=1}^N z_i (1 - x_{1i})} + \right. 
									\\& & \left.  \frac{(1-z) x_{1i}}{\sum_{i=1}^N (1-z_i) x_{1i}} +  \frac{(1-z) (1-x_{1i})}{\sum_{i=1}^N (1-z_i)(1- x_{1i})}  \right). \\
\end{eqnarray*}
When fitting data with sequential offsetted regressions (SOR), we use the functional form for $\wone = \wtwo = [1, x_{1i}, t_{ij}, (t_{ij} - 2)_+]'$. Because $x_{1i}$ is time-invariant, the assumptions in Section~3.2 hold, so every model is fitted using GEE with exchangeable working
correlation structure. When outcome dependent sampling is used, we study naive GEE as well as our proposed approach in order to examine the impact of ignoring the biased study designs. Other working correlation models were tested and the results were not substantially different.


For inverse probability weighting (IPW), observations from each subject are weighted by the inverse of $\pi(z)$ for the AVS design and $\pi(z,x_{1i})$ for the EAVS design.  The models are fit using the R package geeM \citep{mcdaniel2013fast} with an exchangeable working covariance structure.

 Inferential validity and efficiency results are displayed in Table~\ref{poissonsim}. While standard GEE is, as expected, valid under SRS (we do not present the results for SRS here) and ES, it is clearly not valid under AVS and EAVS. The SOR-based GEE approaches
described in this paper are approximately valid under the scenarios we studied, as there is little evidence for bias, and coverage is nearly~$95\%$.

Table \ref{poissonsim} also shows the efficiency of each design relative to a SRS design. 
As expected, ES was more efficient than SRS for all parameters related to $X_{1i}$ 
when $X_{1i}$ was rare, but provided little to no efficiency gains when $\mbox{Pr}(X_{1i}=1)=0.5.$ AVS with SOR analyses
led to at least some efficiency gains over SRS for all parameters; however, when $X_{1i}$ was relatively rare, 
ES was more efficient than AVS for contrasts involving $X_{1i}$. EAVS with SOR analyses tended to be at least as efficient as AVS
in all scenarios studied, and when $\mbox{Pr}(X_{1i}=1)=0.15$, its relative efficiency was $2.15$ and $1.82$
compared to SRS for $\beta_{x_1}$ and $\beta_{tx_1},$ respectively. It was
not optimal for studying $\beta_t$, for which the relative efficiency was
$0.78$. This result is due to relative undersampling of subjects with $x_{1i} = 0$, leading to less efficient estimation
of the intercept at each time point.  In the scenarios studied here, IPW estimation with an AVS or EAVS design led to efficiency gains
that followed a similar pattern, but tended to be smaller than those obtained through SOR.

\begin{table}[h!] \centering \caption{Results from the subject-level simulation} \label{poissonsim} \begin{tabular}{ l l r r r r }  \hline
\multicolumn{1}{c}{Design} & \multicolumn{1}{c}{Estimation} & \multicolumn{1}{c}{$\beta_{0} = -1.4$} & \multicolumn{1}{c}{$\beta_{x_1} = 0.4$}  & \multicolumn{1}{c}{$\beta_{t} = -0.1$} &\multicolumn{1}{c}{$\beta_{t x_1} = 0.1$} \\ \hline

\multicolumn{6}{c }{P($x_{1i} = 1$) = 0.15} \\
ES & GEE &    0 (95) & 0 (95) & 1 (95) & 1 (95) \\
& & [0.59] & [1.70] & [0.59] & [1.46]  \\
AVS & Naive & -39 (0) & -24 (90) & 35 (44) & 4 (94) \\
	& IPW & 0 (95) & 0 (94) & 0 (95) & -1 (94) \\
	&& [1.36] & [1.23] & [1.12] & [1.09] \\ 
	& SOR & 2 (94) & 2 (95) & 0 (95) & -2 (95) \\
& & [1.23] & [1.30] & [1.21] & [1.36]  \\
EAVS & Naive & -42 (0) & -63 (35) & 37 (59) & 14 (91) \\
	& IPW & 0 (95) & -1 (95) & 1 (94) & 1 (94) \\
	&& [0.82] & [2.10] & [0.64] & [1.52] \\
	& SOR & 2 (95) & 2 (95) & -1 (95) & -1 (95) \\
& & [0.78] & [2.15] & [0.76] & [1.82]  \\
 \hline
\multicolumn{6}{c}{P($x_{1i} = 1$) = 0.5} \\
ES & GEE &    0 (96) & 0 (95) & 2 (95) & 2 (95) \\
& & [1.07] & [1.02] & [1.03] & [1.01] \\
AVS & Naive & -35 (0) & -21 (88) & 32 (72) & 4 (94) \\
	& IPW & 0 (96) & -2 (95) & 1 (94) & -1 (94) \\
	&& [1.24] &  [1.26] &  [1.10] & [1.09] \\
	& SOR & 2 (95) & 1 (96) & -1 (94) & -2 (94) \\
& & [1.20] & [1.31] & [1.23] & [1.25]  \\
EAVS & Naive & -41 (0) & -62 (36) & 37 (57) & 14 (92) \\
	& IPW & 0 (95) & 0 (94) & 1 (94) & 1 (95) \\
	&& [1.29] & [1.22] & [1.18] & [1.17] \\
	& SOR & 2 (95) & 4 (95) & -1 (94) & -1 (95) \\ 
& & [1.24] & [1.28] & [1.39] & [1.41]  \\ \hline
\end{tabular}

NOTE: Percent bias, coverage percentage (in parentheses), and empirical efficiency relative to
random sampling (in square brackets) across 2000 replicates of the subject-level sampling Poisson simulation.  Percent bias in parameter estimates is calculated with $100
\cdot ( \widehat{\beta}_k - \beta_k )/\beta_k$ for $ k \in \{ 0, x_1, t, t x_1\}$. Coverage percentages were calculated as the percent of nominal 95\% Wald confidence intervals using standard errors~(\ref{stderrs}) spanning the true parameter value. SRS = simple random sampling;
ES = exposure dependent sampling; AVS= auxiliary variable dependent
sampling; EAVS=exposure and auxiliary variable dependent sampling.
\end{table}

As a means of examining operating characteristics of estimators when sampling probabilities are unknown, we investigated scenarios with sampling ratio ($\pione/\pizero$) misspecification.  IPW and SOR simulations described above were repeated, but when fitting the data, we misspecified the sampling ratio by a factor of 3/2 or 2/3 overall (Table~\ref{misspecify}) and only in those with $x_{1i}=1$ (Table~\ref{stratmiss}).  Results from IPW and SOR were similar.  With overall sampling ratio misspecification, we observed the largest biases in $\beta_0$ and $\beta_t$, which, is consistent with bias in the estimated prevalences at each timepoint.  With sampling ratio misspecification in those with $x_{1i}=1$, we observed biased estimates in contrasts involving $x_{1i}$.  Not surprisingly, time-specific prevalences in those with $x_{1i}=0$ appeared approximately unbiased since their sampling ratio was properly specified.  In general, when sampling probabilities are unknown, we recommend sensitivity analyses to examine the extent to which results are robust to sampling ratio assumptions.


\begin{table}[h!] \centering \caption{Results from the subject-level simulation with misspecified sampling ratios} \label{misspecify} \begin{tabular}{ l l r r r r }  \hline
\multicolumn{1}{c}{Design} & \multicolumn{1}{c}{Estimation} & \multicolumn{1}{c}{$\beta_{0} = -1.4$} & \multicolumn{1}{c}{$\beta_{x_1} = 0.4$}  & \multicolumn{1}{c}{$\beta_{t} = -0.1$} & \multicolumn{1}{c}{$\beta_{t x_1} = 0.1$} \\ \hline

\multicolumn{6}{c}{P($x_{1i} = 1$) = 0.15; $\frac{2}{3} \times   \frac{ \pione}{ \pizero}$} \\
AVS	& IPW & -16 (12) & -8 (94) & 16 (84) & 1 (94) \\
	& SOR & -15 (14) & -8 (94) & 13 (87) & 0 (95) \\
EAVS	& IPW &  -16 (30) &  -9 (94) &  17 (88) & 3 (94) \\
	& SOR & -15 (33) & -7 (94) & 13 (91) & 0 (94) \\
 \hline
\multicolumn{6}{c}{P($x_{1i} = 1$) = 0.5;  $\frac{2}{3} \times \frac{ \pione}{ \pizero}$ }\\
AVS	& IPW & -16 (33) & -9 (94) & 16 (89) & 1 (94) \\
	& SOR & -16 (36) & -9 (94) & 14 (91) & 0 (94) \\
EAVS	& IPW &  -16 (31) & -8 (94) & 17 (89) & 3 (95) \\
	& SOR & -15 (35) & -6 (95) & 13 (91) & 0 (95) \\ \hline 
\multicolumn{6}{c}{P($x_{1i} = 1$) = 0.15; $1.5 \times \frac{ \pione}{ \pizero}$} \\
AVS	& IPW & 16 (22) & 6 (94) & -17 (86) & -3 (93) \\
	& SOR & 17 (17) & 11 (93) & -17 (86) & -7 (95) \\
EAVS	& IPW & 16 (46) & 5 (95) & -16 (89) & -1 (95) \\
	& SOR & 17 (36) & 10 (94) & -19 (86) & -5 (94) \\
 \hline
\multicolumn{6}{c}{P($x_{1i} = 1$) = 0.5;  $1.5 \times \frac{ \pione}{ \pizero}$ }\\
AVS	& IPW & 16 (48) & 4 (96) & -17 (90) & -2 (94) \\
	& SOR & 17 (40) & 9 (95) & -19 (87) & -7 (94) \\
EAVS	& IPW & 16 (45) & 6 (94) & -16 (89) & -1 (95) \\
	& SOR & 17 (35) & 11 (94) & -19 (87) & -5 (94) \\ \hline
\end{tabular}

NOTE: Percent bias and coverage percentage (in parentheses) across 2000 replicates of the subject-level sampling Poisson simulation.  AVS= auxiliary variable dependent
sampling; EAVS=exposure and auxiliary variable dependent sampling.
\end{table}

\begin{table}[h!] \centering \caption{Results from the subject-level simulation with misspecified sampling ratios for subjects with $x_{1i} = 1$.} \label{stratmiss} \begin{tabular}{ l l rrrr }  \hline
\multicolumn{1}{c}{Design} & \multicolumn{1}{c}{Estimation} & \multicolumn{1}{c}{$\beta_{0} = -1.4$} & \multicolumn{1}{c}{$\beta_{x_1} = 0.4$}  & \multicolumn{1}{c}{$\beta_{t} = -0.1$} & \multicolumn{1}{c}{$\beta_{t x_1} = 0.1$} \\ \hline

\multicolumn{6}{c }{P($x_{1i} = 1$) = 0.15; $\frac{2}{3} \times   \frac{ \pione}{ \pizero}$} \\
AVS	& IPW & 0 (95) & 47 (75) & 0 (95) & -14 (92) \\
	& SOR & 1 (95) & 47 (74) & -1 (95) & -14 (92) \\
EAVS	& IPW &  0 (95) &  46 (68) &  1 (94) & -12 (92) \\
	& SOR & 1 (95) & 49 (64) & -2 (95) & -15 (90) \\
 \hline
\multicolumn{6}{c}{P($x_{1i} = 1$) = 0.5;  $\frac{2}{3} \times \frac{ \pione}{ \pizero}$ }\\
AVS	& IPW & 0 (96) & 46 (69) & 1 (94) & -14 (92) \\
	& SOR & 0 (95) & 47 (68) & -2 (94) & -15 (91) \\
EAVS	& IPW &  0 (95) & 48 (68) & 1 (94) & -12 (92) \\
	& SOR & 1 (95) & 50 (64) & -2 (94) & -15 (90) \\ \hline 
\multicolumn{6}{c }{P($x_{1i} = 1$) = 0.15; $1.5 \times \frac{ \pione}{ \pizero}$} \\
AVS	& IPW & 0 (95) & -50 (81) & 0 (95) & 15 (91) \\
	& SOR & 1 (95) & -46 (81) & -1 (95) & 9 (93) \\
EAVS	& IPW & 0 (95) & -51 (66) & 1 (94) & 17 (90) \\
	& SOR & 1 (95) & -47 (68) & -2 (94) & 12 (92) \\
 \hline
\multicolumn{6}{c}{P($x_{1i} = 1$) = 0.5;  $1.5 \times \frac{ \pione}{ \pizero}$ }\\
AVS	& IPW & 0 (96) & -52 (66) & 1 (94) & 15 (92) \\
	& SOR & 0 (96) & -48 (68) & -2 (94) & 10 (93) \\
EAVS	& IPW & 0 (95) & -50 (64) & 1 (94) & 17 (92) \\
	& SOR & 1 (95) & -46 (71) & -2 (94) & 12 (93) \\ \hline
\end{tabular}

NOTE: Percent bias and coverage percentage (in parentheses) across 2000 replicates of the subject-level sampling Poisson simulation.  AVS= auxiliary variable dependent
sampling; EAVS=exposure and auxiliary variable dependent sampling.
\end{table}

\subsection{Observation-Level Sampling}

To investigate the observation-level case, we generated 10 observations for each of 5000 subjects with
\begin{equation*}
Y_{ij} = \beta_0 + \beta_{x_1}x_{1i} + \beta_t t_{ij} + \beta_{x_1, t} x_{1i} t_{ij} + \epsilon_{ij}.
\end{equation*}
Here, $x_{1i}$ was time-invariant and binary, with $P(X_{1i} = 1) = 0.05 \text{ or } 0.1$.  We
also included a time covariate $t_{ij}$ and the time by $x_{1i}$ interaction in the model.  Errors $\epsilon_{ij}$ were 
Gaussian (normal) with an exchangeable correlation structure and parameter~$\alpha=0.3$.

The sampling indicator, $Z_{ij}$, was based on a latent variable $W_{ij} =
\lvert Y_{ij} - \mu + \xi_{ij} \rvert$, where $\xi_{ij}$ is normally distributed, $Z_{ij} = 1_{ \{W_{ij} > \delta \} }$, and $\mu$ is the true mean of the responses.  That sampling is based on an
error-prone version of $Y_{ij}$ reflects a design
wherein sampling is based on an inexpensive and easily obtained surrogate
for a more expensive and accurate response~$Y_{ij}$. Use of the absolute
value function reflects a design wherein extreme values of the response are sampled more often.
The variance of $\xi_{ij}$ is set such that the percentage of the variance of $Y_{ij} + \xi_{ij}$ explained by $Y_{ij}$ 
is 80\%.  The threshold, $\delta$, is set such that 10\% of observations have $Z_{ij} = 1$.  For outcome dependent sampling, observations with $Z_{ij} = 1$ are sampled with certainty, while observations with $Z_{ij} = 0$ are sampled 11\% of the time.
For simple random sampling, observations are sampled at random with a probability so that the expected sample size for simple random sampling is the same as for outcome dependent sampling.

We choose $h(y) = \lvert y \rvert$ in Equation~\ref{sample lambda}, and the linear predictors we use in modeling $\lamyobs$ are 
$\wone = \wtwo = (1, x_{1i}, t_{ij})$. The working correlation structure for observation-level sampling must always be independence, as discussed in Schildcrout et al.\cite{schildcrout2012outcome}. For the GEE analysis when the design uses simple
random sampling, we use exchangeable working covariance weighting which improves efficiency.

Results for the observation-level simulations are shown in
Table~\ref{obssims}.  For simple random sampling, the
percent bias was close to zero and the coverage probability was 95\%, as
expected (results not shown).  When the sampling scheme is ignored, 
the bias was very high for $\beta_{x_1}$.  For SOR and IPW, bias is appropriately corrected, and coverage probability is approximately correct.  Using SOR, we see large efficiency gains for all coefficients.  For IPW, efficiency is 
degraded compared to simple random sampling, which is due to the highly variable weights.  
Many authors have noted that IPW analysis can be unstable and inefficient with highly variable or highly skewed sampling weight distributions.  See Robins et al.\cite{robins1995analysis, robins2000marginal} for a couple of examples.  Schildcrout et al.\cite{Schildcrout2013} observed in simulations (see Table~2) that IPW can, in fact, be less efficient than random sampling.




\begin{table}[h!] \centering \caption{Results from the observation-level simulation} \label{obssims} \begin{tabular}{  l r r r r }  \hline  Estimation & $\beta_0 = 0 $ & $\beta_{x_1} = 0.5 $ & $\beta_{t} = 0 $ & $\beta_{x_1, t} = 0$\\ \hline
 \multicolumn{5}{c}{$R^2_Y = 1.2 \%$, $P(X_{1i}=1) = 0.05$} \\
 Naive & -3 (83) & 106 (1) & 0 (95) & 0 (94) \\ 
SOR & 0 (94) & -1 (94) & 0 (95) & 0 (94) \\
 & [1.87] & [1.66] & [2.17] & [1.87]  \\ 
IPW & 0 (95) & 0 (94) & 0 (96) & 0 (95) \\
 &  [0.79] & [0.77] & [0.77] & [0.72]  \\ \hline
Estimation  & $\beta_0 = 0 $ & $\beta_{x_1} = 0.68 $ & $\beta_{t} = 0 $ & $\beta_{x_1, t} = 0$\\  \hline
 \multicolumn{5}{c}{$R^2_Y = 4\%$, $P(X_{1i}=1) = 0.1$} \\ 
 Naive & -8 (29) & 97 (0) & 0 (94) & 1 (94) \\ 
SOR & 0 (94) & -1 (95) & 0 (94) & 0 (94) \\
 & [1.81] & [1.50] & [2.02] & [1.75]  \\ 
IPW & 0 (95) & 0 (95) & 0 (95) & 0 (94) \\
 &  [0.76] & [0.76] & [0.76] & [0.72]  \\ \hline
 \end{tabular} 

NOTE: Percent bias (or average absolute bias multiplied by 100 when the coefficient is zero), coverage percentage (in parentheses) and efficiency of AVS relative to SRS (in square brackets) across 2000 replicates of the observation-level sampling Gaussian
simulation.
\end{table} 

\section{Applications}

\subsection{ADHD Study}

The ADHD study seeks to identify risk and prognostic factors in early childhood (minimum age at baseline is 3 years, maximum is 7 years) with subjects sampled based on whether ($Z_i =1$) or not ($Z_i = 0$) they were referred to one of two clinics; non-clinical subjects were
sampled from the community. Details are presented elsewhere \citep{lahey1998validity,hartung2002sex,S:Schildcrout+Rathouz:2010}.  Briefly, 25 of the 46 girls were referred and 113 of the 209
boys were referred. The study was matched on gender, $G_i$, so the probability of being sampled depends on the pair $(Z_i, G_i)$. Demographic characteristics were similar between referred and non-referred groups. Subjects were followed for up to 9 years. We assume 5\%\ population
prevalence among girls, $\prob(Z_i | G_i= 1) = 0.05$, so we have $\pi(1,1)/\pi(0,1) = (25*0.95)/(21*0.05) = 22.6$. We assume 15\% prevalence among boys, yielding $\pi(1,0) / \pi(0,0) = 6.7$.

The goal of this analysis is to estimate the time trend of mean
hyperactivity symptom count (analogous to ADHD prevalence, as in Schildcrout and Rathouz\cite{S:Schildcrout+Rathouz:2010}) for boys and girls separately. Using a log-linear Poisson regression model, we fit the response to the covariates given in
Table~\ref{adhdfit}.  The auxiliary model is given by $\wone = \wtwo = (\text{factor}(t_{ij}), \mbox{age}_i, \mbox{female}_i, \mbox{aa}_i, \mbox{other}_i)$ (aa$=$African American;
other$=$other race/ethnicity). The naive
method uses standard GEE. Exchangeable working covariance weighting is
used for all analyses.

The SOR analysis estimates that, at baseline, female gender was associated with $38\%$ fewer ADHD symptoms than male gender (95\% CI: 8\% - 59\%), and
african american race was associated with 49\% (95\% CI: 19\% - 90\%) more ADHD symptoms than white race. It also estimates a significantly different time trend in the number of ADHD symptoms after wave 3 (at $t_{ij}=2$) as compared to prior to wave 3. That is, it estimates a
non-significant upward trend prior to wave 3 (1.06 per wave, 95\% CI: 0.96 - 1.16), that exhibited a significant shift downward with the per wave change in the expected number of ADHD symptoms after wave 3 being 89\% (95\% CI: 80\% - 98\%) of what it was before it. Even though the
naive analysis would also have identified the association with race, neither the gender nor the shift in the time trend would have been detected. Both analyses yielded marginally significant baseline age associations with symptoms.

\begin{table}[h] \centering \caption{Results from the ADHD Study} \label{adhdfit} \begin{tabular}{ l r r r r  } \hline  & \multicolumn{2}{c}{Naive} & \multicolumn{2}{c}{SOR} \\ \hline 
Intercept     & $ 4.76 $ & $( 3.90  , 5.87  )$ & $ 2.89 $ &  $( 2.34    , 3.53  )$ \\
$t$           & $ 0.93 $ & $( 0.86  , 1.02  )$ & $ 1.06 $ &  $( 0.96    , 1.16  )$ \\
$(t-2)_+$     & $ 1.00 $ & $( 0.91  , 1.11  )$ & $ 0.89 $ &  $( 0.80    , 0.98  )$ \\
age           & $ 0.82 $ & $( 0.69  , 0.97  )$ & $ 0.89 $ &  $( 0.76    , 1.02  )$ \\
sex           & $ 0.99 $ & $( 0.62  , 1.58  )$ & $ 0.62 $ &  $( 0.41    , 0.92  )$ \\
afr           & $ 1.32 $ & $( 1.00  , 1.73  )$ & $ 1.49 $ &  $( 1.19    , 1.90  )$ \\
other         & $ 0.88 $ & $( 0.40  , 1.93  )$ & $ 0.87 $ &  $( 0.45    , 1.67  )$ \\
sex*$t$       & $ 1.04 $ & $( 0.82  , 1.32  )$ & $ 1.07 $ &  $( 0.85    , 1.34  )$ \\
sex*$(t-2)_+$ & $ 0.91 $ & $( 0.70  , 1.21  )$ & $ 0.90 $ &  $( 0.70    , 1.15  )$ \\ \hline \end{tabular} 

NOTE: Exponentiated parameter estimates with 95\% confidence intervals for
the ADHD model fit. Confidence intervals containing 1 indicate a
coefficient that is not statistically significant. Time $t=0,\ldots,7$.
\end{table}

\subsection{Biocycle Study}
\begin{table*}[!htp] \small \begin{center}  \caption[]{Results from the BioCycle Study } \label{tab:biocycle} \begin{tabular}{| lcccc |} \hline  & \multicolumn{2}{c}{Estimate (SE)} & \multicolumn{2}{c|}{Geometric mean Ratio (95\% CI)} \\ Variable & SOR & Naive & SOR & Naive
\\ \hline Intercept & 1.425 (0.067) & 1.656 (0.059) & 4.158 (3.644, 4.745) & 5.240 (4.668, 5.883) \\ $Fiber_{ij}$ (per 5 gram/day change) & -0.026 (0.019) & -0.036 (0.019) & 0.974 (0.938, 1.011) & 0.964 (0.928, 1.001) \\ $Calories_{ij}$ (per 1 kcal change) & -0.038 (0.052) &
-0.016 (0.042) & 0.963 (0.870, 1.065) & 0.984 (0.907, 1.067) \\ $BMI_i$ (per 5 kg/m2 change) & -0.075 (0.041) & -0.055 (0.032) & 0.927 (0.857, 1.004) & 0.946 (0.889, 1.007) \\ $Age_i$ (per 10 year change) & -0.080 (0.032) & -0.061 (0.027) & 0.923 (0.867, 0.984) & 0.941 (0.893,
0.992) \\ African American (vs White) & -0.176 (0.073) & -0.213 (0.065) & 0.838 (0.726, 0.967) & 0.808 (0.712, 0.918) \\ Other race (vs White) & -0.109 (0.063) & -0.119 (0.063) & 0.897 (0.793, 1.014) & 0.888 (0.785, 1.006) \\ Cycle 2 (vs. Cycle 1) & 0.048 (0.037) & 0.030 (0.034)
& 1.049 (0.977, 1.127) & 1.030 (0.965, 1.101) \\ \hline \end{tabular} \end{center} \end{table*}

The BioCycle Study  examined relationships among sex hormone levels and
between hormone levels and measures of oxidative stress and dietary intake over the course of the menstrual cycle (Schisterman et al, 2010); its design has been reported
previously (Wactawski-Wende et al., 2009; Howards et al., 2009, Schildcrout et al., 2012). Briefly, on approximately days 6 to 16 of each menstrual cycle, participants used a daily, urine-based fertility test. On the day the monitor indicated peak fertility, i.e., a luteinizing
hormone (LH) surge is highly likely, and the two days following, participants visited the study clinic for peri-ovulation blood draws. We consider fertility
monitor peaks to be a marker for days when LH levels are likely to be high (`high-risk' days; $Z_{ij}=1$) and we assume that on all other days, LH levels are likely to be low (`low-risk' days; $Z_{ij}=0$).

In an earlier report, Schildcrout et al.\cite{schildcrout2012outcome} examined an artificially dichotomized response $I(LH_{ij} \geq 20)$. We examine its relationship with fiber intake on the natural, continuous LH scale. Even though serum was collected on all eight of the key timepoints, fiber
intake was ascertained on only one of the three peri-ovulation visit days, and on three other cycle days (2, 7, and 22 from a standard cycle). Because the design sampled with probability $\sim$ 0.33 (i.e. 1/3) when $Z_{ij}=1$ and with probability $\sim$ 0.12 (i.e., 3/25 in a
typical cycle) when $Z_{ij}=0$, the sample over-represents days with high LH levels.  The skewed LH distribution was log transformed prior to modeling.

In the present analysis, 246 participants admitted data for 450 cycles
with a median age, body mass index, and cycle length of 25 years,
$23\ \mathrm{kg}/\mathrm{m}^2$, and 28 days respectively. The ($5^{th},$
$50^{th}$, $95^{th}$) percentiles of LH and daily dietary fiber intake
were (1.8, 4.8, 25.2) IU/L, and (4.5, 12.2, 29.1) grams per day,
respectively. We evaluated changes in the mean of log-transformed LH
associated with a 5 gram per day (the equivalent of an apple) change in
daily dietary fiber intake. Other covariates included in the regression
model and results from the analysis are shown in Table~\ref{tab:biocycle}. Our auxiliary model used all displayed covariates in
$\wone$ and $\wtwo,$ and the identity for the $h(\cdot).$ Using the SOR
analysis, we estimated that fiber intake was associated with a
$-2.6\%\ (95\% \ CI: \ -6.2\%, \ 1.1\%)$ change in the geometric mean LH
concentrations per five gram per day change in fiber intake. Even though
qualitative conclusions from the naive analysis are not inconsistent with
the SOR analysis, the sample was substantially biased, and we would not
have known how that bias would affect our results had we not undertaken a
bias-correcting approach.  The naive analysis fiber effect estimate was
approximately one-half of a standard error different. Similar sized
differences were also observed for the BMI, age, and African American
gender effect estimates. Overall, we observe that higher baseline age and
African American (versus white) race are associated with lower
geometric mean LH values. A 10-year change in age was associated with a
-7\% (95\% CI: -2, -13)\% change in geometric mean LH and african american
race was associated with a 16\% (95\% CI: 3 - 27\%) lower geometric mean
LH as compared to white race.

\section{Discussion}

 \label{discussion} We have considered analysis for longitudinal data collected under biased outcome- or auxiliary variable-dependent sampling designs, whether at the level of observations or that of subjects, when marginal models are of interest. We have
developed a general GEE-based estimation strategy and robust standard error calculations that incorporate information and uncertainty associated with the study design through an auxiliary model. We also addressed estimation of a dispersion parameter.  An R package, SOR, is available on CRAN.

Through simulation, we have shown that our proposed estimation procedures, if properly specified, are approximately valid and can improve estimation efficiency dramatically over simple random sampling.  
  In contrast, naive analyses of the designs may well result in biased estimators.  Further, since bias was observed for some parameters with sampling ratio misspecification, we recommend sensitivity analyses in circumstances where sampling probabilities ($\pione$ and $\pizero$) are unknown.  

Our main focus here has been on developing a valid approach to data
analysis under biased sampling for longitudinal data. We find that, for binary and count responses, 
efficiency is improved by oversampling at higher values of the outcome distribution.  On the other hand
 it appears to be beneficial to oversample at high and low values of the outcome for continuous responses.

In currently ongoing
research, we are thoroughly characterizing the scenarios under which the
designs are particularly useful in the sense of yielding large efficiency
gains, and when they are unlikely to be useful owing to minimal-to-modest
efficiency gains.

\section{Acknowledgements}

McDaniel's efforts were funded by NIGMS grant 5T32GM074904.  Schildcrout and Rathouz's efforts were funded by NHLBI grant R01 HL094786.
\par

\appendix

\section{Dispersion Parameter Estimation}

 Proceeding from equation~(2.3), we find a score equation and solve it with the assumption of independence between observations.  What follows is for the subject-level sampling setting, but extension to observation-level sampling is straightforward.
\begin{align*} \nonumber
L  & =\prod_{i}{ \prod_{j }{  \frac{\mbox{Pr}(S_i =1 | Y = Y_{ij}, \mb{X}_i ) dF_P (Y_{ij} | \mb{X}_i )}{\int \mbox{Pr}(S_i =1 | Y = y, \mb{X}_i )  dF_P (y | \mb{X}_i ) \times }}} \\ 
& = \prod_{i} { \prod_{j }{  \frac{\exp \left\{ [ \theta_{ij} Y_{ij} - b (\theta_{ij}) ] / \phi + c(Y_{ij}; \phi) + \log \rho_{ij}(Y_{ij}, \mb{X}_i) \right\}}{\int_y \exp \left\{ [ \theta_{ij} y - b (\theta_{ij}) ] / \phi + c(y; \phi) + \log \rho_{ij}(y, \mb{X}_i) \right\}dy }}}
\end{align*}
then taking the log, we get
\begin{align*}
\log L  = & \frac{1}{\phi} \left\{ \sumofsubj \left[ \theta_{ij} Y_{ij} - b (\theta_{ij}) \right] \right\} + \sumofsubj c(Y_{ij}; \phi) + \sumofsubj \log \rho_{ij}(Y_{ij}, \mb{X}_i) \\
& - \sumofsubj \log \int_y\exp \left\{ [ \theta_{ij} y - b (\theta_{ij}) ] / \phi + c(y; \phi) + \log \rho_{ij}(y, \mb{X}_i) \right\}dy.
\end{align*}

\section{Calculation of Standard Errors}

In $\mb{I}_{UT}$, 
\begin{align*}
\frac{\partial \mu_{S_{ij}}}{\partial \gamma_1} = & \wone \frac{\int_y y \times \{ F_{ij}(y_0) - F_{ij}(y) \} \mbox{\odds} (y,\mb{X}_{i} ) dy}{\int_y \mbox{odds}_S (y,\mb{X}_{i} )dy} -  \nonumber \\ 
& - \wone \mu_{S_{ij}} \frac{\int_y   \{ F_{ij}(y_0) - F_{ij}(y) \} \mbox{\odds} (y,\mb{X}_{i} ) dy}{\int_y \mbox{odds}_S (y,\mb{X}_{i} )dy},
\end{align*}
\begin{align*}
\frac{\partial \mu_{S_{ij}}}{\partial \gamma_2} = & \wtwo \frac{\int_y y \times \{ h(y_0) F_{ij}(y_0) - h(y) F_{ij}(y) \} \mbox{\odds} (y,\mb{X}_{i} ) dy}{\int_y \mbox{odds}_S (y,\mb{X}_{i} )dy} -  \nonumber \\ 
& - \wtwo \mu_{S_{ij}} \frac{\int_y   \{ h(y_0) F_{ij}(y_0) - h(y) F_{ij}(y) \} \mbox{\odds} (y,\mb{X}_{i} ) dy}{\int_y \mbox{odds}_S (y,\mb{X}_{i} )dy},
\end{align*}
and
\begin{equation*}
F_{ij}(y) = \lamy(1 - \lamy) \frac{1 - \{ \pione / \pizero \}}{1 - \lamy + \{\pione / \pizero \} \lamy}.
\end{equation*}


\bibliography{refs}

\begin{thebibliography}{10}
\providecommand{\url}[1]{\texttt{#1}}
\providecommand{\urlprefix}{URL }
\expandafter\ifx\csname urlstyle\endcsname\relax
  \providecommand{\doi}[1]{DOI:\discretionary{}{}{}#1}\else
  \providecommand{\doi}{DOI:\discretionary{}{}{}\begingroup
  \urlstyle{rm}\Url}\fi
\providecommand{\eprint}[2][]{\url{#2}}

\bibitem{pmid19159403}
Wactawski-Wende J, Schisterman EF, Hovey KM et~al.
\newblock {BioCycle Study: Design of the Longitudinal Study of the Oxidative
  Stress and Hormone Variation During the Menstrual Cycle}.
\newblock \emph{Paediatric and Perinatal Epidemiology} 2009; 23: 171--184.

\bibitem{pmid18974081}
Howards PP, Schisterman EF, Wactawski-Wende J et~al.
\newblock {Timing Clinic Visits to Phases of the Menstrual Cycle by Using a
  Fertility Monitor: the BioCycle Study}.
\newblock \emph{American Journal of Epidemiology} 2009; 169: 105--112.

\bibitem{schildcrout2012outcome}
Schildcrout JS, Mumford SL, Chen Z et~al.
\newblock {Outcome-Dependent Sampling for Longitudinal Binary Response Data
  Based on a Time-Varying Auxiliary Variable}.
\newblock \emph{Statistics in Medicine} 2012; 31(22): 2441--2456.

\bibitem{lahey1998validity}
Lahey BB, Pelham WE, Stein MA et~al.
\newblock {Validity of {DSM-IV} Attention-Deficit/Hyperactivity Disorder for
  Younger Children}.
\newblock \emph{Journal of the American Academy of Child \& Adolescent
  Psychiatry} 1998; 37(7): 695--702.

\bibitem{hartung2002sex}
Hartung CM, Willcutt EG, Lahey BB et~al.
\newblock {Sex Differences in Young Children Who Meet Criteria for Attention
  Deficit Hyperactivity Disorder}.
\newblock \emph{Journal of Clinical Child and Adolescent Psychology} 2002;
  31(4): 453--464.

\bibitem{S:Schildcrout+Rathouz:2010}
Schildcrout JS and Rathouz PJ.
\newblock {Longitudinal Studies of Binary Response Data Following Case-Control
  and Stratified Case-Control Sampling: Design and Analysis}.
\newblock \emph{Biometrics} 2010; 66(2): 365--373.

\bibitem{BIOM:BIOM12108}
Neuhaus JM, Scott AJ, Wild CJ et~al.
\newblock {Likelihood-Based Analysis of Longitudinal Data From Outcome-Related
  Sampling Designs}.
\newblock \emph{Biometrics} 2014; 70(1): 44--52.

\bibitem{RathouzGao}
Rathouz PJ and Gao L.
\newblock Generalized linear models with unspecified reference distribution.
\newblock \emph{Biostatistics} 2009; 10(2): 205--218.

\bibitem{SIM:SIM557}
Lee AJ, McMurchy L and Scott AJ.
\newblock {Re-using Data From Case-Control Studies}.
\newblock \emph{Statistics in Medicine} 1997; 16(12): 1377--1389.

\bibitem{BIOM:BIOM450}
Neuhaus JM, Scott AJ and Wild CJ.
\newblock {Family-Specific Approaches to the Analysis of Case-Control Family
  Data}.
\newblock \emph{Biometrics} 2006; 62(2): 488--494.

\bibitem{sullivan1994cautionary}
Pepe M and Anderson G.
\newblock {A Cautionary Note on Inference for Marginal Regression Models With
  Longitudinal Data and General Correlated Response Data}.
\newblock \emph{Communications in Statistics-Simulation and Computation} 1994;
  23(4): 939--951.

\bibitem{schildcrout2005regression}
Schildcrout JS and Heagerty PJ.
\newblock {Regression Analysis of Longitudinal Binary Data With Time-Dependent
  Environmental Covariates: Bias and Efficiency}.
\newblock \emph{Biostatistics} 2005; 6(4): 633--652.

\bibitem{liang1992multivariate}
Liang KY, Zeger SL and Qaqish B.
\newblock {Multivariate Regression Analyses for Categorical Data}.
\newblock \emph{Journal of the Royal Statistical Society Series B
  (Methodological)} 1992; : 3--40.

\bibitem{fitzmaurice1995caveat}
Fitzmaurice GM.
\newblock {A Caveat Concerning Independence Estimating Equations With
  Multivariate Binary Data}.
\newblock \emph{Biometrics} 1995; : 309--317.

\bibitem{mancl1996efficiency}
Mancl LA and Leroux BG.
\newblock {Efficiency of Regression Estimates for Clustered Data}.
\newblock \emph{Biometrics} 1996; : 500--511.

\bibitem{robins1994}
Robins JM, Rotnitzky A and Zhao LP.
\newblock Estimation of regression coefficients when some regressors are not
  always observed.
\newblock \emph{Journal of the American Statistical Association} 1994; 89(427):
  846--866.
\newblock \doi{10.1080/01621459.1994.10476818}.
\newblock
  \eprint{http://amstat.tandfonline.com/doi/pdf/10.1080/01621459.1994.10476818}.

\bibitem{robins1995analysis}
Robins JM, Rotnitzky A and Zhao LP.
\newblock Analysis of semiparametric regression models for repeated outcomes in
  the presence of missing data.
\newblock \emph{Journal of the American Statistical Association} 1995; 90(429):
  106--121.

\bibitem{yahav2012}
Yahav I and Shmueli G.
\newblock On generating multivariate poisson data in management science
  applications.
\newblock \emph{Applied Stochastic Models in Business and Industry} 2012;
  28(1): 91--102.
\newblock \doi{10.1002/asmb.901}.

\bibitem{mcdaniel2013fast}
McDaniel LS, Henderson NC and Rathouz PJ.
\newblock Fast pure r implementation of gee: application of the matrix package.
\newblock \emph{The R journal} 2013; 5(1): 181.

\bibitem{robins2000marginal}
Robins JM, Hernan MA and Brumback B.
\newblock Marginal structural models and causal inference in epidemiology.
\newblock \emph{Epidemiology} 2000; : 550--560.

\bibitem{Schildcrout2013}
Schildcrout JS, Garbett SP and Heagerty PJ.
\newblock Outcome vector dependent sampling with longitudinal continuous
  response data: Stratified sampling based on summary statistics.
\newblock \emph{Biometrics} 2013; 69(2): 405--416.

\end{thebibliography}

\end{document}